\documentclass[11pt]{article}

\usepackage{amsmath}
\usepackage{graphicx}
\usepackage{indentfirst}
\usepackage{amssymb}
\usepackage{cite}
\usepackage{color}

\begin{document}

\title{\bf{Stability of Horizon in Warped AdS Black Hole via Particle Absorption}}

\date{}
\maketitle

\begin{center}
\author{Bogeun Gwak}$^a$\footnote{rasenis@sejong.ac.kr}\\

\vskip 0.25in
$^{a}$\it{Department of Physics and Astronomy, Sejong University, Seoul 05006, Republic of Korea}\\
\end{center}
\vskip 0.6in

{\abstract
{We investigate the stability of the horizon in a warped anti-de Sitter black hole based on the three-dimensional new massive gravity by particle absorption. If a particle moving towards the black hole enters its outer horizon, the black hole changes because it absorbs conserved quantities of the particle. This variation is constrained by the equations of motion for the particle. We prove both the irreducibility of the entropy and stability of the horizon for the black hole through this process. However, the instability of the horizon is found in a near-extremal black hole by considering its second-order expansion. We resolve this instability by applying an adiabatic process for the absorption. Further, we show that the stability has the physical counterpart to the energy spectrum of the Virasoro generator via the warped anti-de Sitter/warped conformal field theory correspondence.}
}

\thispagestyle{empty}
\newpage
\setcounter{page}{1}

\section{Introduction}

Anti-de Sitter (AdS) spacetime is a solution to theories of gravity with negative cosmological constant. Physical implications of the AdS spacetime are not easy to find in our Universe by a direct way, because our Universe is estimated to have a positive cosmological constant. However, the gravity theory associated with $N$-dimensional AdS bulk has a holographic correspondence with the conformal field theory (CFT) defined on the $(N-1)$-dimensional boundary of AdS spacetime\cite{Maldacena:1997re,Gubser:1998bc,Witten:1998qj,Aharony:1999ti}.  This is known as AdS/CFT correspondence, and its various applications are related both to quantum chromodynamics (QCD)\cite{Babington:2003vm,Kruczenski:2003uq,Sakai:2005yt,Erlich:2005qh} and condensed matter theory (CMT)\cite{Hartnoll:2008vx,Hartnoll:2008kx}. In the AdS/CFT correspondence, AdS black holes play an important role, because their thermodynamic properties appear in the dual CFT on the AdS boundary\cite{Witten:1998zw}. Then, the dual theory of the AdS black hole corresponds to a finite-temperature CFT, and its temperature is defined as the Hawking temperature in the bulk. In three-dimensional AdS spacetime, entropy of a Ba\~{n}ados-Teitelboim-Zanelli (BTZ) black hole is described by the Cardy formula of the dual CFT based on AdS/CFT correspondence\cite{Banados:1992wn,Banados:1992gq,Cardy:1986ie,Strominger:1997eq}.

Various correspondences have recently been found in non-AdS spacetime. Among these non-AdS holographies, we focus on the duality of a three-dimensional warped Anti-de Sitter (WAdS$_3$) black hole in the bulk. WAdS$_3$ spacetime is AdS$_3$ spacetime deformed by squashing or stretching\cite{Bengtsson:2005zj}. Defined at its WAdS$_3$ boundary, a dual CFT is proposed for the two-dimensional warp conformal field theory (WCFT$_2$). WCFT$_2$ is imposed to provide translation-invariance along with chiral scaling symmetry, such that the extended local algebra consists of the Virasoro algebra plus the U(1) Kac-Moody algebra\cite{Detournay:2012pc}. WAdS$_3$/WCFT$_2$ correspondence is found in both topologically massive gravity (TMG) and new massive gravity (NMG)\cite{Anninos:2008fx,Donnay:2015iia}. TMG is three-dimensional Einstein gravity with a Chern-Simon term\cite{Deser:1981wh,Deser:1982vy} that propagates one polarization of a spin-2 particle. In addition, TMG is a parity-violating theory\cite{Donnay:2015joa}. However, based on this work, the NMG is a parity-preserving model that resolves these issues\cite{Bergshoeff:2009hq}. Similar to four-dimensional gravity, it propagates two polarizations of a spin-2 particle. Therefore, in the NMG, the WAdS$_3$/WCFT$_2$ correspondence is physically close to its four-dimensional counterpart, which is an advantage.

Because the thermodynamics of a black hole in the bulk are related to the dual CFT at the boundary, the stability of the black hole becomes important.  A stable horizon is particularly physically significant, because the black hole's thermal properties, such as its Hawking temperature and Bekenstein-Hawking entropy, are defined on its horizon\cite{Christodoulou:1970wf,Christodoulou:1972kt,Bekenstein:1973ur,Bekenstein:1974ax,Hawking:1974sw,Hawking:1976de}. The stability of the horizon is primarily studied because of the cosmic censorship conjecture, in which the singularity of a black hole is assumed to be covered by its horizon\cite{Penrose:1969pc}. The validity of this conjecture can be investigated by testing whether a black hole can be overspun or not. This test has been performed on various black holes in a number of gravity theories\cite{Wald:1974ge,Dadhich:1997rq,Hubeny:1998ga,Jacobson:2009kt,Barausse:2011vx,Colleoni:2015afa,Hod:2016hqx,Natario:2016bay,Horowitz:2016ezu,Gwak:2015fsa,Duztas:2016xfg,Gwak:2016gwj,Sorce:2017dst}. As a black hole can either obtain or lose rotational energy via the Penrose process\cite{Bardeen:1970zz,Penrose:1971uk}, we can investigate its stability under particle absorption. In this case, the laws of thermodynamics is strongly related to the stability of the horizon\cite{Gwak:2015fsa,Gwak:2016gwj}.

In this work, we investigate the stability of the horizon in the WAdS$_3$ black hole through particle absorption. The WAdS$_3$ black hole is a solution to the NMG, which is a ghost-free theory propagating two polarizations of a spin-2 particle. Based on these considerations, we can realize a similar physics in three-dimensional spacetime to that of four dimensions. Furthermore, a description of the holography in three dimensions is obtained, and WAdS$_3$/WCFT$_2$ correspondence, which was studied thoroughly in \cite{Donnay:2015iia}, is considered. By doing so, we obtain not only the changes in the WAdS$_3$ black hole under particle absorption but also its physical implications with respect to dual WCFT$_2$. In particular, by testing if the black hole can be overspun by particle absorption, we demonstrate that the horizon remains stable under this process, and this stability is explained from the perspective of dual WCFT$_2$. In addition, we prove the instability of the horizon for the first- and second-order expansions of the near-extremal black hole. However, this is resolved when a physical absorption process is assumed. Although a WAdS$_3$ black hole is regular, the stability of its horizon plays an important role in applications of WAdS$_3$/WCFT$_2$ correspondence, because its thermal properties are defined on its horizon.

This paper is organized as follows. In Sec.\,\ref{sec2}, WAdS$_3$/CFT$_2$ is outlined briefly. In Sec.\,\ref{sec3}, the changes in a WAdS$_3$ black hole are obtained in terms of the conserved quantities of the particles absorbed. In Sec.\,\ref{sec4}, the stability of an extremal WAdS$_3$ black hole horizon is investigated under particle absorption. In Sec.\,\ref{sec5}, the horizon of a near-extremal black hole is tested and demonstrated to be stable under the adiabatic processes. In Sec.\,\ref{sec6}, the stability of the horizon that we have found is explained in terms of a dual description of WCFT$_2$. Finally, in Sec.\,\ref{sec7}, our results are summarized.

\section{WAdS$_3$/WCFT$_2$ Correspondence in New Massive Gravity}\label{sec2}

The WAdS$_3$ black hole represents a solution of the NMG that is a ghost-free theory with two degrees of freedom in propagation. The action of NMG, which is a parity-even theory, is defined by the high-curvature term\cite{Bergshoeff:2009hq}
\begin{align}\label{eq:action1}
S=\frac{1}{16\pi}\int d^3 x \sqrt{-g} \left(R-2\Lambda +\frac{1}{m^2}\left(R_{\mu\nu} R^{\mu\nu}-\frac{3}{8}R^2\right)\right)\,,
\end{align}
where $m$ and $\Lambda$ are the mass of a graviton and the cosmological constant, respectively. The equations of motion are obtained from Eq.\,(\ref{eq:action1}) as
\begin{align}\label{eq:eom2}
G_{\mu\nu}+\Lambda g_{\mu\nu} +\frac{1}{2m^2}K_{\mu\nu}=0\,,
\end{align}
where
\begin{align}
G_{\mu\nu}&=R_{\mu\nu}-\frac{1}{2}Rg_{\mu\nu}\,,\nonumber\\
K_{\mu\nu}&=2\square R_{\mu\nu} - \frac{1}{2}\nabla_\mu \nabla_\nu R -\frac{1}{2}\square R g_{\mu\nu} + 4 R_{\mu\alpha\nu\beta}R^{\alpha \beta} -\frac{3}{2}R R_{\mu\nu}-R_{\alpha\beta}R^{\alpha\beta}g_{\mu\nu} +\frac{3}{8}R^2 g_{\mu\nu}\,.\nonumber
\end{align}
WAdS$_3$ spacetimes are solutions to Eq.\,(\ref{eq:eom2})\cite{Bengtsson:2005zj,Clement:2009gq} and are correlated to the dual WCFT$_2$. This correspondence is represented as non-AdS holography\cite{Anninos:2008fx,Song:2011sr,Detournay:2012pc}. Entropy of a WAdS$_3$ black hole corresponds to the Cardy formula from the dual WCFT$_2$ provided in \cite{Donnay:2015iia}. The WAdS$_3$ spacetimes include the timelike and spacelike solutions that play important roles in WAdS$_3$/WCFT$_2$ correspondence. The metric of the timelike WAdS$_3$ spacetime is the same as that of the G\"{o}del spacetime, which is given as\cite{Banados:2005da}
\begin{align}\label{eq:godelsp}
ds^2 = -dt^2 -4 \omega r dt d\phi +\frac{\ell^2 dr^2}{(2r^2(\omega^2 \ell^2 +1)+2\ell^2 r)}-\left(\frac{2r^2}{\ell^2}(\omega^2 \ell^2 -1)-2r\right)d\phi^2\,,
\end{align}  
and is a solution to Eq.\,(\ref{eq:eom1}) in terms of the parameter choices
\begin{align}
m^2 = - \frac{19\omega^2 \ell^2 -2}{2\ell^2}\,,\quad \Lambda = -\frac{11\omega^4 \ell^4+28\omega^2 \ell^2-4}{2\ell^2(19\omega^2\ell^2-2)}\,.
\end{align}
As the limit of $\omega^2 \ell^2 $ approaches unity, the solution of Eq.\,(\ref{eq:godelsp}) approaches AdS$_3$ spacetime. In G\"{o}del spacetime, the physical mass is read in NMG\cite{Donnay:2015joa} as
\begin{align}
\mathcal{M}_{\text{G}}=-\frac{4\ell^2\omega^2}{G(19\ell^2\omega^2-2)}\,,
\end{align}
which will be associated to the vacuum of the WAdS$_3$ black hole introduced as follows.

A WAdS$_3$ black hole is a spacelike solution to Eq.\,(\ref{eq:eom1}) that preserves the isometry group SL(2, $\mathbb{R}$)$\times$U(1)\cite{Clement:2009gq,Moussa:2003fc}. This solution includes the parameter $\nu$, called the warp factor, which is associated to the asymptotic geometry of the spacetime. The black hole represents asymptotically stretched AdS$_3$ when $\nu>1$ and asymptotically squashed AdS$_3$ when $\nu<1$. When $\nu=1$, its asymptotic geometry is AdS$_3$ spacetime. In this work, we focus on a squashed spacelike black hole for which the warp factor is always larger than unity. In terms of ADM formalism, a WAdS$_3$ black hole is given by\cite{Anninos:2008fx,Donnay:2015iia}
\begin{align}\label{eq:metric1}
ds^2& = -N(r)^2 dt^2+\ell^2 R(r)^2(d\phi+N^\phi(r)dt)^2 +\frac{\ell^4 dr^2}{4R(r)^2 N(r)^2}\,,
\end{align}
where 
\begin{align}
R(r)^2&= \frac{r}{4}\left(3(\nu^2-1)r+(\nu^2+3)(r_++r_-)-4\nu\sqrt{r_+r_-(\nu^2+3)}\right)\,,\nonumber\\
N(r)^2&=\frac{\ell^2(\nu^2+3)(r-r_+)(r-r_-)}{4R(r)^2}\,,\quad N^{\phi}=\frac{2\nu r - \sqrt{r_+r_-(\nu^2+3)}}{2R(r)^2}\,.\nonumber
\end{align}
Note that $g_{tt}=-N(r)^2+\ell^2 R(r)^2(N^{\phi})^2=\ell^2$, so this represents a spacelike spacetime. The inner and outer horizons are represented by $r_-$ and $r_+$, respectively, and $\ell$ is a free parameter. Note also that one can consider the behaviors of a particle in spacelike spacetime as though it were in the ergosphere of a Kerr black hole\cite{Clement:2009gq}. This particle's motion cannot be stable throughout the entire spacelike spacetime. The equations of motion are solved as in the choices of parameters
\begin{align}
m^2 = -\frac{20\mu^2 -3}{2\ell^2}\,,\quad \Lambda = - \frac{m^2(4\nu^2 - 48\nu^2 +9)}{400\nu^4-120\nu^2 +9}\,.
\end{align}  
The mass and angular momentum of the black hole are given as\cite{Clement:2009gq,Giribet:2012rr,Nam:2010ub}
\begin{align}\label{eq:defmj03}
\mathcal{M}&=\frac{\nu(\nu^2+3)}{G\ell (20\nu^2-3)}\left((r_+ + r_-)\nu-\sqrt{r_+ r_- (\nu^2+3)}\right)\,,\\
\mathcal{J}&=\frac{\nu(\nu^2+3)}{4G\ell (20\nu^2-3)}\left((5\nu^2+3)r_+ r_- -2 \nu (r_+ + r_-)\sqrt{r_+ r_- (\nu^2+3)}\right)\,.\nonumber
\end{align}
The inner and outer horizons of an extremal black hole are coincident with one another, meaning that $r_-=r_+$. Hence, the extremal condition is given as
\begin{align}\label{eq:extremalcon1}
\mathcal{J} \leq \frac{G\ell(20\nu^2-3)}{4\nu(\nu^2+3)}\mathcal{M}^2\,.
\end{align}
The entropy of the black hole is obtained using the Wald formula\cite{Wald:1993nt,Giribet:2012rr}
\begin{align}\label{eq:entropyw}
S_{\text{BH}}&= \frac{8\pi \nu^3}{(20\nu^2-3)}\left(r_+ - \frac{1}{2\nu}\sqrt{(\nu^2+3)r_+ r_-}\right)\,.
\end{align}
To identify the Cardy formula for the dual WCFT$_2$, the set of boundary conditions\cite{Compere:2009zj}
\begin{align}
g_{tt}&=\ell^2 + \mathcal{O}(r^{-1})\,,\quad \quad\quad\quad\,\,\,\,\, g_{tr}=\mathcal{O}(r^{-2})\,,\quad  \,g_{t\phi}=\ell^2 \nu r+\mathcal{O}(1)\,,\\
g_{rr}&=\frac{\ell^2}{(\nu^2+3)r^2} + \mathcal{O}(r^{-3})\,,\quad g_{r\phi}=\mathcal{O}(r^{-1})\,,\quad  g_{\phi\phi}=\frac{3}{4}r^2\ell^2(\nu^2-1)+\mathcal{O}(r)\,,\nonumber
\end{align}
which are imposed and used to obtain the generators of diffeomorphism. Using these generators, the charge algebra is equivalent to the semi-direct sum of Virasoro algebra with the central charge $c$ and the affine $\hat{u}(1)_k$ Ka\v{c}-Moody algebra of level $k$ and normalization $N_1$. Then, two commuting Virasoro algebras generated by $L_n^+$ and $L_m^-$ are defined. The eigenvalues of the energy spectra $L^\pm_0$ are provided in terms of the WAdS$_3$ black hole configuration
\begin{align}
h^+ = \frac{1}{k}\mathcal{M}^2-\mathcal{J}-\frac{c}{24}\,, \quad h^- = \frac{1}{k}\left(\mathcal{M}+\frac{kN_1}{2}\right)^2\,,
\end{align}
where
\begin{align}
k=\frac{4\nu(\nu^2+3)}{G\ell(20\nu^2-3)}\,,\quad N_1=\frac{2\ell \nu}{\nu^2+3}\,,\quad c=-\frac{96\ell\nu^3}{G(20\nu^4+57\nu^2-9)}=-6kN_1^2\,.\nonumber
\end{align}
The eigenvalue of the energy spectrum $L_0^{\pm}$ is bounded because of the extremal condition in Eq.\,(\ref{eq:extremalcon1}). Therefore,
\begin{align}
h^\pm\geq -\frac{c}{24}\,.
\end{align}

The entropy of the WAdS$_3$ black hole in Eq.\,(\ref{eq:entropyw}) corresponds to the Cardy formula for dual WCFT$_2$, which represents asymptotic state growth. This correspondence is proven in terms of the shifted Virasoro operators $\tilde{L}_0^\pm$ with the eigenvalue of the energy spectrum $\tilde{h}^\pm$ to be
\begin{align}
h^+\rightarrow \tilde{h}^+=\frac{1}{k}\mathcal{M}^2-\mathcal{J}\,,\quad h^-\rightarrow \tilde{h}^-=\frac{1}{k}\mathcal{M}^2\,,
\end{align}
for which the Cardy formula is written as
\begin{align}
S_{\text{CFT}}=2\pi \sqrt{-4\tilde{h}^{-\text{(vac)}}\tilde{h}^-}+2\pi\sqrt{-4\tilde{h}^{+\text{(vac)}}\tilde{h}^+}\,,
\end{align}
where the values of $\tilde{h}^{\pm\text{(vac)}}$ read in the vacuum geometry proposed for the G\"{o}del geometry\cite{Detournay:2012pc,Donnay:2015iia}. Hence,
\begin{align}
\tilde{h}^{\pm\text{(vac)}}=\frac{1}{k}(\mathcal{M}^{\text{(vac)}})^2\,,\quad \mathcal{M}^{\text{(vac)}}=i\mathcal{M}_{\text{G}}=-i\frac{4\ell^2\omega^2}{G(19\ell^2\omega^2-2)}\,,\quad \omega^2\ell^2=\frac{2\nu^2}{3-\nu^2}\,.
\end{align} 
Thus, the Cardy formula for WCFT$_2$ corresponds exactly to the entropy of the AdS$_3$ black hole\cite{Donnay:2015iia}.
\begin{align}
S_{\text{CFT}}=S_{\text{BH}}\,.
\end{align}

\section{Particle Absorption in WAdS$_3$ Black Hole}\label{sec3}

To change a black hole, we need to change its conserved quantities. These changes should be transferred to the black hole in physically defined ways, such as via particle absorption. When a particle enters a black hole, its conserved quantities merge and vary those of the black hole, according to the particle's equations of motion. We investigate how black hole change through physical processes under particle absorption and relate these changes to the dual WCFT$_2$ theory, finding that the changes can be obtained in terms of the momenta carried by the particles. The Hamiltonian and Hamilton-Jacobi actions for a particle with a mass $\mu$ are given by
\begin{align}\label{eq:actions1}
\mathcal{H}=\frac{1}{2}g^{\mu\nu}p_\mu p_\nu\,, \quad S=\frac{1}{2}\mu^2\lambda+Et+L\phi+S_r(r)\,,
\end{align}
where the momentum is defined as $p_\mu=\partial S / \partial x^\mu$, and the affine parameter is $\lambda$. Since the metric in Eq.\,(\ref{eq:metric1}) has translation symmetries for coordinates $t$ and $\phi$, the corresponding conserved quantities are defined as $E$ and $L$ in the Hamilton-Jacobi action. In addition, the mass of the particle is given as $m$. We consider a particle moving under a spacelike metric of Eq.\,(\ref{eq:metric1}); thus, positive values are chosen for $E$ in Eq.\,(\ref{eq:actions1}) because $g_{tt}$ in Eq.\,(\ref{eq:metric1}) is positive. The translation symmetry of coordinate $t$ still generates the conserved quantities $E$. However, $t$ is now spacelike in Eq.~(\ref{eq:metric1}), similar to $r$.  Further, in the Hamilton-Jacobi equation
\begin{align}\label{eq:eom1}
-\frac{\partial S}{\partial \lambda}=-\mu^2=-\frac{1}{N(r)^2}(E-N^\phi L)^2+\frac{L^2}{R(r)^2 \ell^2}+\frac{4N(r)^2R^2(r)}{\ell^4}(\partial_r S(r))^2\,,
\end{align}
where the sign in front of $\mu^2$ is still negative. This is similar to a particle in the ergosphere of a Kerr black hole due to the metric component $g_{tt}$. The sign in from of $\mu^2$ remains negative, indicating that the motion of the particle in the ergosphere is unstable. From the same perspective, $\mu^2$ is still negative in Eq.\,(\ref{eq:eom1}), so the motion of a particle is unstable throughout the entire spacetime generated by the WAdS$_3$ black hole. However, the negative sign in front of $\mu^2$ is removed as $r$ approaches the limit of the outer horizon $r_+$ in the equations of motion, implying that it does not affect the following results. It is assumed that the particle is absorbed by the black hole when it passes through the outer horizon; therefore, the relationships among the conserved quantities need to be obtained at the outer horizon. As $r$ approaches a limit of $r_{+}$, the Hamilton-Jacobi equation in Eq.\,(\ref{eq:eom1}) is reduced to
\begin{align}\label{eq:particleenergy1}
E=N^{\phi}_{+}L \pm \frac{\ell^2 |p^r|}{2R_{+}}\,,
\end{align}
where the radial momentum of the particle is defined as $p^r\equiv\dot{r}=\partial r/\partial \lambda$ at the outer horizon, and
\begin{align}\label{eq:parameter02}
R_{+}^2&\equiv R^2(r_{+}) = \frac{r_+}{4}\left(2\nu\sqrt{r_+}-\sqrt{(\nu^2+3)r_-}\right)^2\,,\\
N^\phi_{+}&\equiv N^\phi(r_{+}) = \frac{2}{\sqrt{r_+}(2\nu\sqrt{r_+}-\sqrt{(\nu^2+3)r_-})}\,.\nonumber
\end{align}
Although the coordinate $t$ is spacelike, it is still assumed that $E$ represents the energy of the particle in Eq.\,(\ref{eq:particleenergy1}). Therefore, the energy should be positive with respect to the infalling particle as time increases; thus, we make this term positive in Eq.\,(\ref{eq:particleenergy1}). This choice ensures that the entropy of the black hole increases, which complies with the second law of thermodynamics.

We can now obtain the relationships among the conserved quantities of the particle passing through the outer horizon. At the moment, the particle enters the horizon. Thus, it can no longer be observed from outside of the horizon. Hence, we assume that the black hole has absorbed the particle, and its conserved quantities are considered to be part of the WAdS$_3$ black hole. This means that the conserved quantities of the black hole vary infinitesimally based on the particle's quantities. Thus, we assume that
\begin{align}\label{eq:variation01}
E=dM\,,\quad L=dJ\,,
\end{align}
which complies both with the conservation of mass and the conservation of angular momentum. Based on Eq.\,(\ref{eq:variation01}), the variations in the mass and the angular momentum are related to one  another by Eqs.\,(\ref{eq:particleenergy1}) and (\ref{eq:variation01}) with positive sign chosen, meaning that
\begin{align}\label{eq:mastereq00}
dM=N^{\phi}_{+}dJ + \frac{\ell^2 |p^r|}{2R_{+}}\,.
\end{align}
Because the metric of Eq.\,(\ref{eq:metric1}) is provided in terms of $r_+$ and $r_-$, the variations of $dM$ and $dJ$ can be rewritten as $dr_+$ and $dr_-$, respectively, from Eq.\,(\ref{eq:defmj03}),
\begin{align}\label{eq:variationmj02}
d M =&\frac{\nu(\nu^2+3)}{G\ell (20 \nu^2-3)}\left(\nu-\frac{r_- (\nu^2+3)}{2\sqrt{r_+ r_- (\nu^2+3)}}\right)d r_+ + \frac{\nu(\nu^2+3)}{G\ell (20 \nu^2-3)}\left(\nu-\frac{r_+ (\nu^2+3)}{2\sqrt{r_+ r_- (\nu^2+3)}}\right)d r_-\,,\nonumber\\
d J =& \frac{\nu(\nu^2+3)}{4G\ell (20\nu^2-3)}\left(-\frac{r_- (r_+ + r_-)\nu(\nu^2+3)}{\sqrt{r_+ r_-(\nu^2+3)}}-2\nu\sqrt{r_+ r_- (\nu^2+3)}+r_-(5\nu^2+3)\right)d r_+\\
&+ \frac{\nu(\nu^2+3)}{4G\ell (20\nu^2-3)}\left(-\frac{r_+ (r_+ + r_-)\nu(\nu^2+3)}{\sqrt{r_+ r_-(\nu^2+3)}}-2\nu\sqrt{r_+ r_- (\nu^2+3)}+r_+(5\nu^2+3)\right)d r_-\,.\nonumber
\end{align}
Solving Eq.\,(\ref{eq:variationmj02}) with Eq.\,(\ref{eq:mastereq00}), we obtain the variation in $r_+$ as a complex form in terms of $L$ and $p^r$. This shows that the change in the outer horizon depends on both the radial and angular momenta of the particle. Therefore, the change is indirect. However, instead of the outer horizon, the entropy of the black hole is affected by particle absorption, given by
\begin{align}
d S_{BH}=&\frac{8\pi \nu^3}{G(20\nu^2-3)}\left(1-\frac{r_- (\nu^2+3)}{4\nu\sqrt{r_+ r_- (\nu^2+3)}}\right)d r_+ -\frac{2\pi r_+ \nu^2 (\nu^2+3)}{G(20\nu^2-3)\sqrt{r_+ r_- (\nu^2+3)}}d r_-\,,
\end{align}
in which both $dr_+$ and $dr_-$ can be written in terms of the particle's momenta by inserting Eqs.\,(\ref{eq:mastereq00}) and (\ref{eq:variationmj02}). Then, the change in entropy is reduced to a very simple form as
\begin{align}\label{eq:entropy01}
d S_{BH}= \frac{4\pi \ell^3 |p^r|}{(\nu^2+3)(r_+ - r_-)}> 0\,.
\end{align}
This implies that the increase in entropy under particle absorption can be obtained. In addition, the irreducible property can be identified by particle absorption. This differs from the case of the BTZ black hole in which the irreducible property is the area of the outer horizon in consideration of AdS/CFT correspondence\cite{Gwak:2012hq}.

It can be demonstrated that the outer horizon remains stable under particle absorption with the increase in entropy defined by the horizon in Eq.\,(\ref{eq:entropy01}). However, the change in entropy is singular at $r_+=r_-$, the extremal condition. Therefore, the stability of the horizon should be tested for the extremal WAdS$_3$ black hole as described in the following section.

\section{Stability of Horizon in Extremal WAdS$_3$ Black Hole}\label{sec4}

An extremal WAdS$_3$ black hole has the maximally saturated angular momentum for a given mass; therefore, it satisfies the equality in Eq.\,(\ref{eq:extremalcon1}). For an extremal black hole, the outer and inner horizons are coincident with one another, and changes in the outer horizon itself are important. Because an infinitesimal increase in the angular momentum leads to a violation of the inequality of the extremal condition in Eq.\,(\ref{eq:extremalcon1}), it is possible that such overcharging disappears the horizon. In this case, the horizon of the extremal black hole is unstable under particle absorption, and the inverse of the metric component $g^{rr}$ at the horizon $r_\text{e}\equiv r_+=r_-$ is written as
\begin{align}
g^{rr}{\big |}_{r=r_\text{e}}=0\,,\quad \partial_r g^{rr}{\big |}_{r=r_\text{e}}=0\,,\quad (\partial_r)^2 g^{rr}{\big |}_{r=r_\text{e}}=\frac{2(\nu^2+3)}{\ell^2}>0\,,
\end{align}
which implies that the horizon of the extremal black hole is located at the minimum point of the component $g^{rr}$. Then, if the particle absorption changes the minimum value of $g^{rr}$ to a positive value, there is no solution for $g^{rr}=0$, so the horizon disappears in the spacetime. On the other hand, if the minimum value becomes negative, the horizon still covers the black hole. However, the change in $g^{rr}$ written in terms of $dr_+$ and $dr_-$ given as Eq.\,(\ref{eq:variationmj02}) cannot be investigated because the existence of a horizon cannot be ensured under absorption. Therefore, the configuration of the black hole after the particle absorption should be obtained by using $dM$ and $dJ$. Instead of testing $g^{rr}$, the extremal condition in Eq.\,(\ref{eq:extremalcon1}) can be tested to verify the final state of the black hole. An extremal black hole $(M,\,J)$ is at the state saturating the equality in Eq.\,(\ref{eq:extremalcon1}). After absorbing a particle, the black hole transitions to the state $(M+dM,\,J+dJ)$. The extremal condition in the final state changes to
\begin{align}\label{eq:extremalcon2}
J^2-\frac{G\ell(20\nu^2-3)}{4\nu(\nu^2+3)}M^2=0\,\rightarrow\, (J+dJ)^2-\frac{G\ell(20\nu^2-3)}{4\nu(\nu^2+3)}(M+dM)^2\,,
\end{align}
where $dM$ is removed by Eq.\,(\ref{eq:mastereq00}) and all parameters are rewritten as $r_e$ using Eqs.\,(\ref{eq:defmj03}) and (\ref{eq:parameter02}). Then, we obtain the exact relation
\begin{align}\label{eq:extremalconfinal}
dJ-\frac{G\ell(20\nu^2-3)}{2\nu(\nu^2+3)}MdM=-\frac{\ell^2 |p^r|}{2}< 0\,,
\end{align}
in which the minimum value always becomes negative under absorption. Because there is no dependency on the angular momentum of the particle in Eq.\,(\ref{eq:extremalconfinal}), the extremal black hole cannot be overspun by absorption. Therefore, the outer horizon remains stable during the process.

We now know that the horizon exists after absorption; therefore, changes to an extremal black hole can be written in terms of $dr_+$ and $dr_-$. Through the component $g^{rr}$, the stability of the horizon can also be identified by particle absorption. The absorption changes the locations of both the outer and inner horizons as well as the minimum point, such that the change in the minimum value of $g^{rr}$ can be written as
\begin{align}\label{eq:grrvariation01}
d g^{rr}(r_e)=\frac{\partial g^{rr}(r_e)}{\partial r_+}d r_+ +\frac{\partial g^{rr}(r_e)}{\partial r_-}d r_- +\frac{\partial g^{rr}(r_e)}{\partial r_e}d r_e\,.
\end{align}
The varied minimum point is no longer located on either horizon. In addition, the outer and inner horizons are no longer coincident after absorption. Inserting Eqs.\,(\ref{eq:mastereq00}) and (\ref{eq:variationmj02}), we can obtain Eq.\,(\ref{eq:grrvariation01}) in terms of the momentum of the particle as
\begin{align}
d g^{rr}(r_e)=\frac{G\ell(-20\nu^2+3)|p^r|}{2\nu^3}<0\,,
\end{align}
which implies that the minimum value of $g^{rr}$ becomes negative. Then, the component $g^{rr}$ has two solutions that correspond to the inner and outer horizons. Thus, the outer horizon is not only stable and existent, but the black hole also becomes non-extremal. Furthermore, this demonstrates that the increase in the black hole mass is always greater than that in its angular momentum.

\section{Stability of Horizon in Near-Extremal WAdS$_3$ Black Hole}\label{sec5}

We demonstrated the stability of the horizon when particle absorption changes the black hole infinitesimally, such that the variation is considered to be the first order of expansion. However, we ignored the second-order terms. If the particle absorption changes the near-extremal black hole slightly more than infinitesimally, it is possible that the black hole will be overspun beyond the near-extremal condition. We now investigate whether this case can be realized under absorption or not. If it is, in fact, possible, we will investigate how to resolve the instability of the horizon. The near-extremal condition is expressed as the near-extremal parameter $D$, a free parameter from Eq.\,(\ref{eq:extremalcon1}), as
\begin{align}
\frac{G\ell(20\nu^2-3)}{4\nu(\nu^2+3)}M^2 -J= D>0\,,
\end{align}
where $D\ll 1$ because of the near-extremal condition. When the parameter $D$ becomes negative under absorption, the black hole is overspun and its horizons disappear in the spacetime. We assume that the initial black hole $(M,\,J)$ becomes $(M+\Delta M,\,J+\Delta J)$ for the final black hole. The energy and angular momentum of the particle, written as $(\Delta M, \,\Delta J)$, are both transferred to the black hole. Instead of $(d M,\,d J)$, the changes in the black hole are indicated by $(\Delta M, \,\Delta J)$, where the variation in $\Delta$ is slightly greater than that of $d$, meaning that the second-order term should be taken account in the expansion. The near-extremal condition is obtained in terms of the near-extremal parameter $D_\Delta$ that is changed by the absorption. Then,
\begin{align}
\frac{G\ell(20\nu^2-3)}{4\nu(\nu^2+3)}(M+\Delta M)^2 -(J+\Delta J)= D_\Delta\,,
\end{align}
which is rewritten with the insertion of Eq.\,(\ref{eq:mastereq00}), giving
\begin{align}\label{eq:secondfulleq1}
D_\Delta=&D-\left(1-\frac{G\ell M N^{\phi}_+ (20\nu^2-3)}{2\nu(\nu^2+3)}\right)\Delta J+\frac{G\ell (N^{\phi}_+)^2 (20\nu^2-3)}{4\nu(\nu^2+3)}(\Delta J)^2+\frac{G\ell^3 N^{\phi} (20\nu^2-3)}{4R_+\nu(\nu^2+3)}\Delta J |p^r|\nonumber\\
&+\frac{G\ell^3 M (20\nu^2-3)}{4R_+\nu(\nu^2+3)}|p^r|+\frac{G\ell^5 (20\nu^2-3)}{16R_+^2\nu(\nu^2+3)}|p^r|^2\,.
\end{align}
For the black hole to be overspun in its final state, the near-extremal parameter $D_\Delta$ must be negative. Hence, negative contributors to $D_\Delta$ , such as the first-order term of $\Delta J$ , play important roles. Note that the first-order term of $\Delta J$ differs significantly from those of previous sections because we are focusing on $\Delta J>dJ$ cases. Thus, if we choose proper values for both $D$ and $\Delta J$, it is possible for the black hole to be overspun. However, we first need to test whether the near-extremal black hole can spin beyond its extremal bound or not. The particle should have only angular momentum $L=\Delta J$ and no radial momentum, or $|p^r|$=0. By considering only the first-order term $\Delta J$, the final near-extremal parameter can be rewritten as an inequality from Eq.\,(\ref{eq:secondfulleq1}). Then, when the near-extremal black hole is overspun by the particle, the range of the angular momentum is a solution of the inequality
\begin{align}\label{eq:inequalityfirstorder}
\Delta J > \frac{D}{F(M,J)}\,,\quad F(M,\,J)=1-\frac{G\ell M N^{\phi}_+ (20\nu^2-3)}{2\nu(\nu^2+3)}\,,
\end{align}
where the parameter $D_{\Delta}$ is negative and $F(M,\,J)$ is positive. Thus, the horizon of a near-extremal black hole is unstable. In addition, this instability can occur when the momentum of the particle is absorbed into the black hole without delay and spreads over its entire volume. In reality, both the energy and the angular momentum of the particle may take a finite period of time to spread throughout the black hole, so the process is dynamic. To consider this effect, we assume an adiabatic process consisting of a number of steps, during which the conserved quantities of the particle transfer to those of the black hole\cite{Chirco:2010rq,Gwak:2016gwj}. Because $\Delta J$ is greater than $dJ$, it should be divided into $N$ distinct $dJ$. Therefore, $\Delta J \equiv NdJ$. Afterwards, we consider $N$ steps for which the black hole absorbs $dJ$ in one of the steps. During this process, the near-extremal parameter $D$ changes to a different $D_{\Delta}$ over the $N$ steps, resulting in
\begin{align}
D\,\rightarrow \,D_{d_1} \,\rightarrow\, D_{d_2}\,\rightarrow \,...\,\rightarrow\, D_{d_{n-1}}\,\rightarrow\,D_{\Delta}\,,
\end{align}
where it is assumed that $\Delta J$ satisfies the inequality of Eq.\,(\ref{eq:inequalityfirstorder}) and $D_{d_n}$ is the near-extremal parameter of the black hole absorbing the angular momentum $ndJ$ over $n$ steps. In addition, the black hole $(M,J)$ changes to $(M_n,J_n)$ at the $n$th step. Each value of $D_{d_n}$ is obtained from the first-order expansion of $\Delta J$ in Eq.\,(\ref{eq:secondfulleq1}) so that
\begin{align}
D_{d_1}=D-F(M,J)\frac{\Delta J}{N}\,,\, D_{d_2}=D_{d_1}-F(M_1,J_1)\frac{\Delta J}{N}\,,\,...\,,\,D_{\Delta}=D_{d_{n-1}}-F(M_{n-1},J_{n-1})\frac{\Delta J}{N}\,,
\end{align}
for which we can choose a value of $N$ that is sufficiently large to make all values of $D_{n}$ positive. Thus, we can expect that the black hole approaches an extremal black hole under continuous absorption in the adiabatic process. However, the black hole is still under the extremal condition, therefore, it cannot be overspun by particle absorption. Using a similar approach, we can determine the instability of the near-extremal black hole with the second-order expansion from Eq.\,(\ref{eq:secondfulleq1}) such that
\begin{align}\label{eq:secondordereq3}
D-F(M,J)\Delta J+K(M,J)(\Delta J)^2<0\,,\quad K(M,J)=\frac{G\ell (N^{\phi}_+)^2 (20\nu^2-3)}{4\nu(\nu^2+3)}>0\,,
\end{align}
for which the range of the solution is
\begin{align}\label{eq:secondordereq2}
\frac{F(M,J)-\sqrt{F(M,J)^2-4DK(M,J)}}{2K(M,J)}<\Delta J <\frac{F(M,J)+\sqrt{F(M,J)^2-4DK(M,J)}}{2K(M,J)}\,.
\end{align}
The solution of $\Delta J$ will exist under $F(M,J)^2-4DK(M,J)>0$ in the choice of proper parameters. Then, for $\Delta J$ which satisfies Eq.\,(\ref{eq:secondordereq2}), each value of $D_{d_n}$ is obtained from Eq.\,(\ref{eq:secondordereq3}) as
\begin{align}
D_{d_1}&=D-F(M,J)\frac{\Delta J}{N}+K(M,J)\left(\frac{\Delta J}{N}\right)^2\,,\\
D_{d_2}&=D_{d_1}-F(M_1,J_1)\frac{\Delta J}{N}+K(M_1,J_1)\left(\frac{\Delta J}{N}\right)^2\,,\nonumber\\
&\quad\quad\quad\quad\quad\quad\quad\quad\quad{\vdots}\nonumber\\
D_{\Delta}&=D_{d_{n-1}}-F(M_{n-1},J_{n-1})\frac{\Delta J}{N}+K(M_{n-1},J_{n-1})\left(\frac{\Delta J}{N}\right)^2\,,\nonumber
\end{align}
for which a value of $N$ large enough to make $D_n$ positive can always be chosen. Therefore, the instability of the horizon can be resolved in the second order expansion under continuous absorption. Even if the particle has an energy and an angular momentum sufficient to overspin the black hole, because the process is physical, the black hole cannot jump over the extremal condition but rather approaches extremality. This is also consistent with the third law of thermodynamics in that there is no physical process that can achieve extremality.

\section{Stability in WAdS$_3$/WCFT$_2$ Correspondence}\label{sec6}

According to WAdS$_3$/WCFT$_2$ duality, there are physical correspondences between the properties of a black hole and dual WCFT$_2$. As reviewed in Sec.\,\ref{sec2}, the entropy of a black hole corresponds to the Cardy formula in dual WCFT$_2$. Thus, we investigate if changes in appear in dual WCFT$_2$ for a the WAdS$_3$ black hole perturbed by a particle in the bulk. Because the Cardy formula is provided in terms of the energy spectrum $\tilde{h}^\pm$, particle absorption can affect the energy spectrum of dual WCFT$_2$.

When combined with the extremal condition in Eq.\,(\ref{eq:extremalcon1}), the energy spectrum $\tilde{h}^\pm$ of the shifted Virasoro operator $\tilde{L}_0^\pm$ is bounded to
\begin{align}
\tilde{h}^\pm\geq 0\,.
\end{align}
However, the bounded points are different for each $\tilde{h}^\pm$. For the energy spectrum $\tilde{h}^+$, the bounded value is saturated to zero in the bulk of the extremal black hole. Then, the change in $\tilde{h}^+$ under particle absorption is obtained as
\begin{align}
d\tilde{h}^+ =\left(-1+\frac{(20\nu^2-3)G\ell M N^\phi}{2\nu(\nu^2+3)}\right)L+\frac{G\ell^3 M(20\nu^2-3)}{4\nu(\nu^2+3)R_+}|p^r|\,,
\end{align}
which is the similar to the change of the extremal condition in Eq.\,(\ref{eq:extremalconfinal}). The energy spectrum $\tilde{h}^+$ increases at the point saturated to the extremal condition due to the sign, so
\begin{align}
d\tilde{h}^+= \frac{\ell^2 |p^r|}{2}\,.
\end{align}
Therefore, the spectrum $\tilde{h}^+$ is ensured to be positive by the stability of the horizon in Eq.\,(\ref{eq:extremalconfinal}).  For the second order of expansion, the stability of the horizon is already ensured by continuous absorption in the adiabatic process. For the spectrum $\tilde{h}^-$, the bounded value is saturated to zero when the mass of the black hole is zero. Its change can be clearly identified in the second-order expansion. Under particle absorption, the mass of the black hole $M$ changes to $M+\Delta M$. Thus, 
\begin{align}
\tilde{h}^- +\Delta \tilde{h}^- = \frac{1}{k}\left(M+\Delta M\right)^2=\frac{1}{k}\left(M+ N^\phi \Delta J +\frac{\ell^2 |p^r|}{2R}\right)^2\geq 0\,.
\end{align}
The spectrum $\tilde{h}^-$ is bounded to zero when $M$ approaches a limit of zero. At this point, the change is obtained as
\begin{align}
\Delta \tilde{h}^- =\frac{1}{k}\left(N^\phi \Delta J +\frac{\ell^2 |p^r|}{2R}\right)^2>0\,.
\end{align}
Thus, the energy spectrum $\tilde{h}^-$ increases for all values of the particle's momentum, because the black hole's mass is positive. Therefore, the mass of the black hole that cannot vanish because of particle absorption is equivalent to the positive energy spectrum $\tilde{h}^-$.

The change in the Cardy formula is written in terms of the energy spectrum under particle absorption as
\begin{align}\label{eq:dualent01}
dS_{CFT}=\frac{2\pi i}{\sqrt{k}} M^{\text{vac}}\left(\frac{d\tilde{h}^+}{\sqrt{\tilde{h}^+}}+\frac{d\tilde{h}^-}{\sqrt{\tilde{h}^-}}\right)= \frac{4\pi \ell^3 |p^r|}{(\nu^2+3)(r_+ - r_-)}=d S_{BH}\,.
\end{align}
Changes in the energy spectrum $\tilde{h}^\pm$ also change the Cardy formula for the dual WCFT$_2$ because of the perturbance of the WAdS$_3$ black hole. Then, the change in the Cardy formula exactly corresponds to that of the entropy of the black hole in Eq.\,(\ref{eq:dualent01}). Therefore, the correspondence between the WAdS$_3$ black hole and dual WCFT$_2$ is still well defined for the first order of expansion.

\section{Summary}\label{sec7}

We investigated the horizon stability of a WAdS$_3$ black hole under particle absorption. According to the WAdS$_3$/WCFT correspondence, this stability correlates with the energy spectrum of the shifted Virasoro generators for the minimum possible energy. To illustrate how the change in the black hole is related to dual WCFT$_2$ under particle absorption, we obtained the equations of motion of a particle absorbed into the black hole considering spacelike geometry. The particle is then assumed to be absorbed, when it passes through the outer horizon of the black hole. The conserved quantities of the particle are expected to be transferred to those of the black hole. Thus, because of these conserved quantities, we posited that the black hole changes infinitesimally and proportionally to their magnitudes. Under absorption, we demonstrated that the entropy of the black hole obtained using Wald formula is irreducible. To test the stability of the horizon, we focused primarily on whether extremal and near-extremal black holes could be overspun by particles or not. However, the increase in an extremal black hole's mass is always greater than that of the angular momentum transferred from the particle. Thus, the horizon is stable under absorption. In the case of the near-extremal black hole, we considered the second-order term of the expansion. When the near-extremal parameter is slightly smaller than the change caused by absorption, a near-extremal black hole can overspin. However, as the absorption takes time to spread its transferred conserved quantities over the entire volume of the black hole, we defined continuous absorption based on an adiabatic process. Consequently, a near-extremal black can approach an extremal black hole more closely than before but cannot be overspun. Therefore, we demonstrated the stability of the horizon under absorption, which allowed us to safely estimate the correlation between the properties of the WAdS$_3$ black hole and dual WCFT. A black hole can be defined in terms of the ground energy spectrum of the shifted WCFT Virasoro generator, and the minimum energies of the spectra are bounded by both the extremality of the black hole and its positive mass. Then, we demonstrated the change in the Cardy formula corresponding to the changes of the energy spectrum caused by particle absorption. Based on this process, the change was exactly identified and matched to that of the first-order expansion of the entropy of the black hole.

\vspace{10pt} 

{\bf Acknowledgments}

This work was supported by Basic Science Research Program through the National Research Foundation of Korea (NRF) funded by the Ministry of Science, ICT \& Future Planning (NRF-2015R1C1A1A02037523).

\end{document}